# Unconventional Computing based on Four Wave Mixing in Highly Nonlinear Waveguides

Kostas Sozos, Stavros Deligiannidis, Charis Mesaritakis, Adonis Bogris

*Abstract*— **In this work we numerically analyze a photonic unconventional accelerator based on the four-wave mixing effect in highly nonlinear waveguides. The proposed scheme can act as a fully analogue system for nonlinear signal processing directly in the optical domain. By exploiting the rich Kerr-induced nonlinearities, multiple nonlinear transformations of an input signal can be generated and used for solving complex nonlinear tasks. We first evaluate the performance of our scheme in the Santa-Fe chaotic time-series prediction. The true power of this processor is revealed in the all-optical nonlinearity compensation in an optical communication scenario where we provide results superior to those offered by strong machine learning algorithms with reduced power consumption and computational complexity. Finally, we showcase how the FWM module can be used as a reconfigurable nonlinear activation module being capable of reproducing characteristic functions such as sigmoid or rectified linear unit.**

## I. INTRODUCTION

NOWADAYS, photonic neuromorphic/unconventional computing has arisen in the form of powerful hardware platforms that tackle difficult tasks and accelerate the processing speed with marginal power consumption [1]. Neuromorphic computing generally consists of linear and nonlinear operations in order to process signals and extract useful patterns from them. Linear transformations are straightforward and efficient in the optical domain. This effectiveness has stimulated an explosion in the investigation and production of linear photonic processors which execute trillions of operations per second by just inferring light through waveguide meshes equipped with phase shifters [2]. Moreover, the inherent parallelization of light permits wavelength multiplexing and increase the available degrees of freedom in the implementation of the synapses of a photonic neural network. On the other hand, nonlinear transformations operating as activation functions on neuromorphic networks fall behind in efficiency, despite being of paramount importance. They are based either on electro-optic solutions, such as electro-absorption modulators [3] and the square law of the photodiode [4], or on purely optical ones which exploit saturable absorbers [5] or Kerr effect [6].

Four-wave mixing (FWM) is one of the principal nonlinear effects in optics. It has found multiple applications, spanning from optical [7] and quantum communications [8] to optical computing, implementing optical subtraction or time-spectral convolution [9]. Here, we propose FWM process as an elegant[1] solution for the generation of multiple, rich and diverse nonlinear products in the complex domain. These products can act as activation functions in different types of neural networks, such as recurrent neural networks (RNNs), extreme learning machines (ELMs) and reservoir computers (RCs) or in any other type of brain-inspired framework. In this work we directly drive FWM products in a digital linear regression stage, following an ELM approach [10], where output layer weights are calculated in order to solve a nonlinear task. Hence, the FWM engine formulates a modular optical nonlinear activation function with programmable nonlinearity depending on the input power and the wavelengths launched into the nonlinear medium. We simulate and benchmark this system in two tasks, first, in the Santa Fe chaotic time series prediction derived from laser-generated data. Then, we evaluate the FWM-processor in the significant, application-wise, mitigation of self-phase modulation in optical communications. Here, we unveil the true power of the system, as both amplitude and phase information play crucial role in the signal equalization, therefore the complex nonlinear activation function offered by the nonlinear coherent process is fully capitalized. In this second task, we compare our system with complex digital equalizers based on machine learning in order to illustrate the merits of the nonlinear computing offered by FWM. Lastly, we showcase how a linear combination of the different products can be leveraged in order to construct typical non-linear activation functions used in machine learning at a very good approximation. Hence, FWM can provide solutions in order to fill the gap of all-optical nonlinear processing in next generation photonic neural networks. In the next section we describe the basic aspects of the idea and we analyze the numerical modelling approach. In section III, we provide the numerical results in our two benchmark tasks and we also analyze the behavior of the proposed modular optical nonlinearity. Finally, in section IV we conclude our work.

Manuscript received XXXX. This work was supported in part by the Hellenic Foundation for Research and Innovation (H.F.R.I.) through the 2nd Call for H.F.R.I. Kostas Sozos, Stavros Deligiannidis and Adonis Bogris are with the Department of Informatics and Computer Engineering, University of West Attica, Aghiou Spiridonos, Egaleo, 12243, Athens, Greece (e-mail: ksozos@uniwa.gr, abogris@uniwa.gr).

Charis Mesaritakis is with the Department of Information & Communication Systems Engineering, University of the Aegean, 2 Palama & Gorgyras St., 83200, Karlovassi Samos, Greece (e-mail: cmesar@aegean.gr)

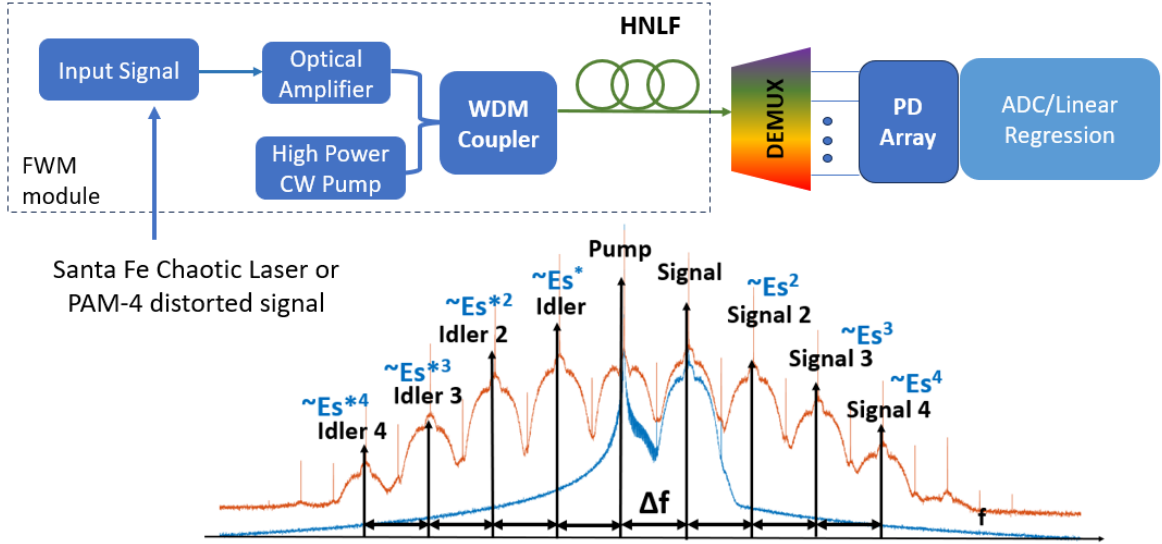

Figure 1. The setup of the proposed system. The signal, the pump and the products that are created through FWM, separated to signals and idlers, are shown in the frequency axis. The input signals that are used in this work are time-series from the well-known Santa Fe Laser dataset and a non-linear transmission scenario.

## II. CONCEPT AND NUMERICAL MODELLING

The neuromorphic system that is proposed here is depicted in Fig. 1. Before being launched into the FWM module, the signal passes through an optical amplifier in order to boost its power. The amplified signal is combined with a high-power continuous wave (CW) pump into a highly nonlinear fiber (HNLF) constituting the FWM apparatus. Integrated nonlinear waveguides can be employed for more compact solutions [11], [12]. The interaction of the amplified input signal with frequency $f_{signal}$, with the CW pump at a frequency $f_{pump}$ in the HNLF will create a first order FWM product in a frequency $f_{idler}$ given by (1) [13].

$$f_{idler} = 2f_{pump} - f_{signal} \quad (1)$$

This process is the so-called degenerate four wave mixing, as only two waves interact instead of three. If the input power is high enough to make the nonlinear process evolve in the saturation regime, a cascaded generation of higher order products will take place resulting in the creation of multiple nonlinear copies of the input signal. The number and the power of these copies is a function of the input power and the dispersion properties of the nonlinear medium which affects phase matching conditions [14]. Fig. 1 shows that the FWM products are either higher order copies of the signal or its conjugate counterpart (idler). In reality, the use of a high-power input signal results in a more complex nonlinear transformation as the pump is also depleted through the process. An optical demultiplexer drives each product to an optical filter followed by a photodiode and an analog-to-digital converter (ADC). The detected signals are weighted and summed in a linear regression stage in the digital domain, implementing a versatile activation function. Additionally, a relative delay between the different products can be introduced in the analog or the digital domain, in order to resemble a fractionally spaced equalizer, that enhances the performance in telecom tasks.

The simulation parameters that are used in this work, for the HNLF and the other types of fibers, are provided in Table I. Moreover, all types of noise have been included in the simulation. Lasers with initial relative intensity noise of -160 dB/Hz were numerically simulated. For both pump and signal amplification we have considered optical amplifiers with noise figure of 5 dB. The thermal and shot noise at the photodetection stage were also taken into account. Finally, the digitization stage was simulated with ADCs operating at baud rate with 8-bit resolution. The nonlinear process was simulated by solving nonlinear Schrodinger equation in the HNLF, using the split step Fourier method with the propagating field including both signal and pump, considering that all dispersion parameters are calculated with respect to pump wavelength [15]. For the HNLF propagation, up to fourth order dispersion is taken into account. Polarization-mode dispersion effects in HLNF have been neglected in order to simplify the simulation process. Pump wavelength is 1551 nm in all cases studied, thus the parametric process takes place in the anomalous dispersion regime and close to zero dispersion wavelength in order to maximize the efficiency and bandwidth of the parametric process [16]. The next section provides the results for different tasks starting with the Santa Fe chaotic laser prediction task.

TABLE I
Numerical Model Parameters

| Parameter | Value | Units |
|---|---|---|
| HNLF Length | 1 | km |
| HNLF Losses | 0.8 | dB km$^{-1}$ |
| HNLF Nonlinear Factor | 12 | W$^{-1}$km$^{-1}$ |
| HNLF Dispersion Slope | 0.03 | ps nm$^{-2}$ km$^{-1}$ |
| HNLF Zero disp. wavelength $\lambda_0$ | 1550 | nm |
| SMF Losses | 0.21 | dB km$^{-1}$ |
| SMF Nonlinear Factor | 1.3 | W$^{-1}$km$^{-1}$ |
| SMF Dispersion Coefficient | 16 | ps nm$^{-1}$ km$^{-1}$ |
| DCF Losses | 0.45 | dB km$^{-1}$ |
| DCF Nonlinear Factor | 5.4 | W$^{-1}$km$^{-1}$ |
| DCF Dispersion Coefficient | -200 | ps nm$^{-1}$ km$^{-1}$ |

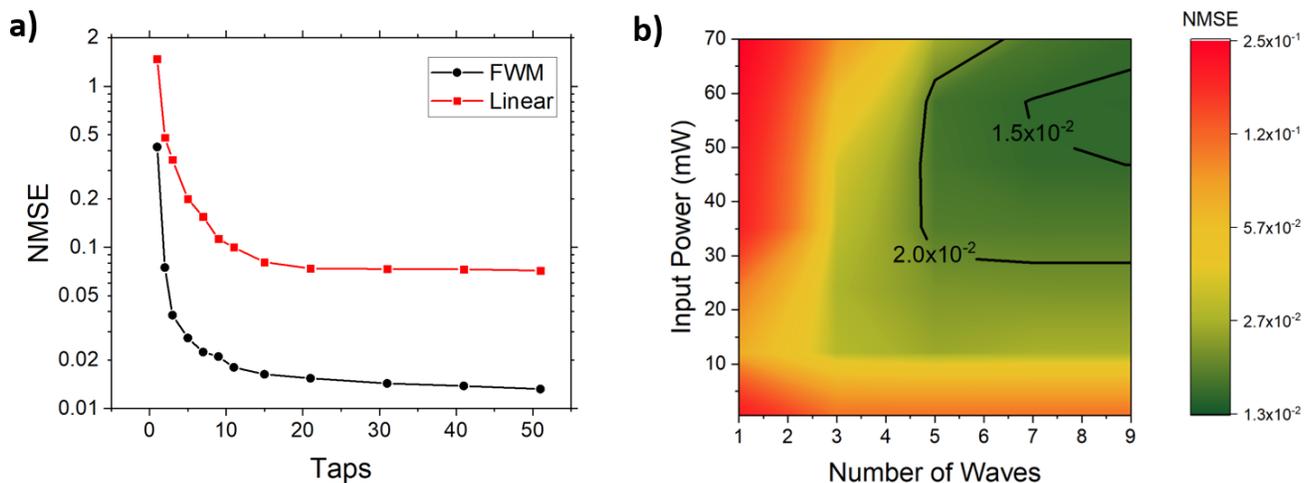

Figure 2. a) Performance optimization vs the number of taps in the Santa-Fe task. A linear algorithm is also investigated for comparison. b) Contour plot of the NMSE versus the input signal power and the number of waves that are kept in the linear regression stage for the Santa Fe task.

## III. RESULTS AND DISCUSSION

### A. Santa Fe time series prediction

Regarding the first task, we numerically modulate a transmitter with 10093 points of the laser generated Santa Fe data at a 20 Gbaud symbol rate. We directly amplify the transmitter output in order to acquire a high-power input signal and then mix it with a CW pump of 200 mW. The pump exhibits Optical Signal to Noise Ratio (OSNR) of 55 dB while the signal has OSNR=45 dB at 0.1 nm. The pump-signal frequency detuning is $\Delta f$=40 GHz in this task. After collecting the outputs from the photodiodes, we perform linear regression with the $k$ past symbols of each output in order to predict the following one, as this task dictates. We choose $k$=11 symbols after an optimization process, as shown in Fig 2a, especially when the number of products taken into account exceeds 5. In this way, we acquire 11x$N$ predictors, where $N$ is the number of wavelengths (including the initial signal) that we provide to the linear regression. We evaluate the performance via the Normalized Mean Square Error (NMSE). Increasing the input signal power for a given HNLF length we can vary the number of newly generated products up to 16 (including both signal and idler products).

In Fig. 2b, we cross-investigate how input power and the number of products that we keep for regression affect the accuracy for the Santa-Fe task. The number of products that can provide a useful set of signals to the regression stage depends on the pump and signal power levels, the nonlinear parameter of the fibre and its length. We keep the pump power at 200 mW. We vary the signal power from 1 to 70 mW and the number of waves that are driven to the regression stage from one (only signal) to 9, spanning from signal to signal 9. In this task we select only signal products (not idler ones), as they are red-shifted and travel slower than the signal itself in the HNLF. Since, in Santa-Fe task it is important to predict the evolution of the time series, it is important to guarantee that dispersion has not assisted the prediction capabilities of the nonlinear system, hence selecting signal products in the regression process reassures that all timeseries are retarded with respect to the initial signal. It must be noted that the considered HNLF has a small dispersion in order to enhance the parametric interaction. The maximum delay is small and estimated to be 0.15 ps between signal and signal 9, thus it does not affect the regression performance. We observe that there is an optimum input power around 55 mW that contributes to the production of at least 8 products with adequate power (>0.1 mW). In Fig. 2b, it is shown that we achieve NMSE performance below $2 \times 10^{-2}$ with 5 or more waves. This is an achievement equivalent to that of contemporary photonic neuromorphic systems which utilize around 400 nodes [17]. Instead, here we only employ 55 nodes (5 physical outputs X 11 taps), which means that FWM phenomenon under saturation is a remarkably nonlinear dynamical system capable of chaotic time series prediction. The performance saturates at $1.3 \times 10^{-2}$ with 8 waves and subsequently, 88 nodes. For comparison, fig. 2a includes the performance versus the number of taps for a purely linear algorithm. The linear system saturates completely at 21 taps, offering NMSE=0.073. For comparison, a FWM system with 20 parameters, i.e 5 waves with 4 taps for each one, would result in NMSE=0.0268, or 270 % improvement.

### B. Non-linear optical channel equalization

We also apply the FWM-based processing to a task related to optical communications, that is the mitigation of fiber nonlinearities. A 50 Gbaud quaternary pulse amplitude modulated (PAM-4) signal is numerically transmitted through a dispersive link consisting of a 50 km long Single-Mode Fiber (SMF) and the chromatic dispersion (CD) is compensated using a 4 km long dispersion compensating fiber (DCF) with its parameters shown in Table I. By increasing the launched power, Kerr-induced nonlinearity arises. Interacting with noise, this nonlinearity heavily distorts the signal and deteriorates the link performance. We take the DCF output signal and we amplify it at 30 mW in order to acquire 4 signal and 4 idler products. The OSNR is adjusted to 45 dB. The amplified output is mixed with a pump of 200 mW, with pump-signal spacing being equal to $\Delta f$=100 GHz. The launched power in the SMF-DCF transmission link affecting the accumulated nonlinear distortion is varied from 12 to 16 dBm.

Here, we consider an external memory of 15 taps for each product, as the optimization process of the bit-error rate (BER)

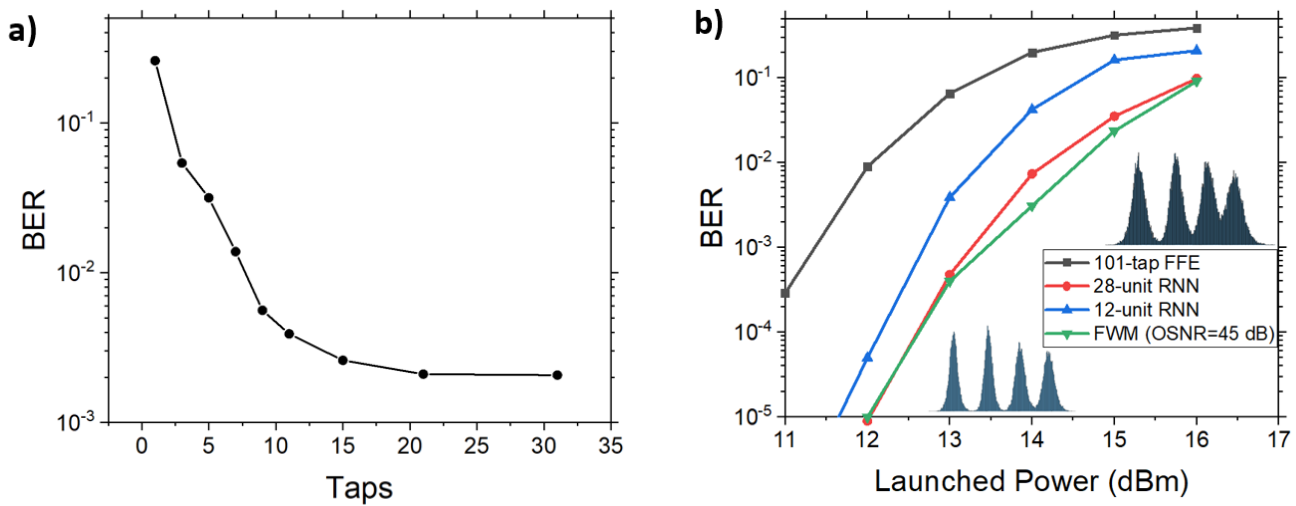

Figure 3. a) Optimization of the performance vs the number of taps for the FWM processor in Kerr Non-Linearities mitigation. b) Performance comparisons in the Kerr Non-Linearities compensation task. BER versus Launched Power for FFE, RNN and FWM processor.

shown in fig. 3a indicates. With the total amount of CD compensated, nonlinear effects are the dominant impairment. Increasing the launched power beyond 12 dBm, harsh nonlinear conditions are present and the linear feed-forward equalizer (FFE) cannot efficiently compensate the distortion, as shown in Fig. 3b even though 101 taps are considered. We provide the performance of the FWM processor with 10 channels (the four signals and four idler products, along with the pump and signal). Therefore, we utilize idler 1 to idler 4 and signal 2 to signal 5 along with the pump and the signal. In this task, we also include a small temporal delay between the products which is a fraction of the symbol period, in order to mimic a fractionally spaced equalizer that enhances the performance in telecom tasks. More specifically, the signal 2 and the idler 1 are delayed by the duration of half a symbol, the signal 3 and the idler 2 by 3/4 of the symbol duration and lastly, signal 4 and idler 3 by 1/4 of the symbol duration. These delays were selected through an exhaustive BER optimization process as well. For comparison, we utilize a nonlinear equalizer in the form of a Recurrent Neural Network (RNN) with either 12 or 28 hidden units and OSNR=45 dB. More specifically, we employ bi-directional vanilla RNNs, exploiting the many-to-many training approach [18]. The RNN models are built, trained and evaluated in Keras with Tensorflow 2.3 GPU backend. In the Keras model, binary cross-entropy is chosen as a loss function and Adam as the optimizer. We considered a single input in the RNN model, that of the transmitted signal. We also consider 80.000 symbols for training, 20.000 for validation and 60.000 for testing. The models and the training of our benchmark RNN equalizers are extensively discussed in [18]. The FWM processor achieves similar BER results with the strong 28-unit RNN thanks to the highly complex nonlinear transformations from the HNLF assisted by the weighted mixing of products. As for the complexity reduction, our proposition employs 150 real multiplications for the 15 taps of each of the 10 waves. On the contrary, our benchmarked RNNs require 638 and 2610 multiplications for the case of 12 and 28 hidden units respectively. It is worth mentioning that although the signal is further nonlinearly distorted by the HNLF propagation, the FWM processing of the distorted signal excels in mitigating the overall nonlinear distortion.

### C. Modular Non-linear activation function

An aspect of great interest for photonic computing community is the generation of non-linear activation functions in an ultra-fast manner and directly in the optical domain. FWM is a quasi-instantaneous phenomenon, offering a mechanism of unparalleled speed [7]. Here, in order to create a specific function, we combine the different nonlinear products, each of them exhibiting a different nonlinear transfer characteristic with respect to the input signal. In Fig. 4a, the input-output transfer characteristics for each of the first five products are shown (Signal 2,3,4 and Idler 1,2), along with the initial signal nonlinear transformation (Signal) and pump depletion which is a result of the saturated FWM process. The nonlinear dependence of the products on the input power and the variety of slopes is obvious. Utilizing this pallet of nonlinear responses, one can approximate a wide range of different activation functions by simply properly weighting the outputs. In Fig. 4b, different typical activation functions are formulated by multiplying each of the seven waves, with a proper weight, that has been estimated through a simple least mean square error algorithm. The NMSE of the calculated activation function from the targeted one lies in the order of $10^{-3}$, for Rectified Linear Unit (ReLU) and sigmoid functions with small steepness, while approaches $10^{-2}$ for the sigmoid function with the higher steepness. The aforementioned results indicate that, through the multiple nonlinear products created in the FWM process, a modular optical activation function of high versatility can be approximated.

### D. Discussion

In this work, highly nonlinear fibers with lengths of hundreds of meters are considered. However, there is a fast progress in highly nonlinear integrated waveguides in various platforms [11], exhibiting nonlinear parameter $\gamma$ over 500 $W^{-1}m^{-1}$. So, in order to achieve up to 5th order terms with pump power of 200 mW and signal power of 30-50 mW, as in this work, 1-2 cm long waveguides could be employed. In terms of system complexity and power consumption, the proposed scheme requires an amplifier and a high-power CW pump that increase the power budget of a system, while in its most compact form would employ a specialized optical circuit with an integrated nonlinear element followed by a demultiplexer and a photodiode array. On the other hand, it vastly simplifies digital signal processing, as it depends only on a linear stage with few multiplications and summations. The major restraint comes from the photodetection and the analog-to digital conversion in order to perform the linear regression in the digital domain. Nevertheless, this could be avoided if silicon photonic microring weight banks are considered after the demultiplexer, as all FWM products reside in different wavelengths, therefore they are compatible with incoherent wavelength-division multiplexed (WDM) neural networks [19]. At that simplified one-photodiode version, that system could offer a very practical and strong nonlinear accelerator, especially for applications dealing with intrinsic optical signals.

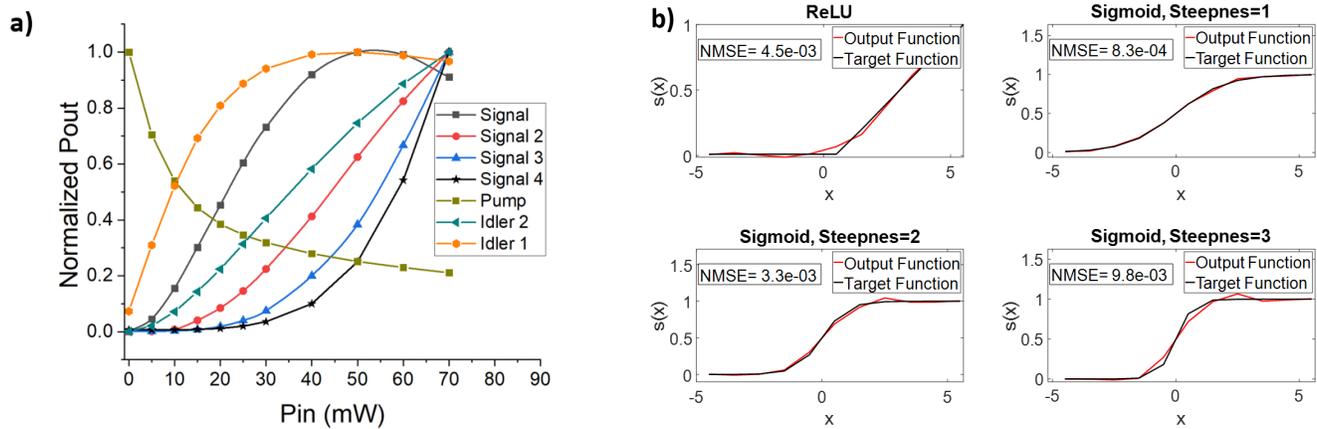

Figure 4. a) The nonlinear input-output characteristics of each of the FWM products. b) Some of the nonlinear activation functions that can be approximated by the set of 7 products multiplied by the proper weight. Even steep sigmoid functions can be produced with acceptable fidelity.

For an indicative power consumption comparison, we will analyze the example of the task of the section IIB. A digital algorithm that runs at 50 Gbaud, would consume over 4 mW/multiplication [20]. So, employing 600 multiplications required by the light RNN version we benchmarked would consume over 2.4 W only for the digital signal processing, imposing latency in the order of μs. On the contrary, a possible one PD-ADC system, would only consume ~450 mW for digital multiplications. Adding the power consumption of the amplifier and the ~200 mW of the high-power pump, this system can tackle a difficult non-linear telecom task, with over 65% energy savings.

IV. CONCLUSION

In this work, we proposed and numerically simulated a photonic neuromorphic processor based on the FWM effect in a HNLF. This FWM-based processor was benchmarked in the Santa Fe time-series prediction task, achieving competitive results compared to state-of-the-art photonic extreme learning machines/reservoir computers with minimum NMSE=0.014. It was also benchmarked in the mitigation of Kerr-induced transmission impairments in a 50 Gbaud PAM-4 optical link. In this task, the FWM processor outperformed a strong 28-unit bi-RNN equalizer. Lastly, a modular activation function was demonstrated, achieving ReLU and sigmoid behavior. The aforementioned results show that FWM can enable a wide range of nonlinear transformations being useful for many neuromorphic computing/machine learning paradigms. Further investigations on this scheme will include evaluation in more complex tasks, masking techniques at the pump wavelength in order to unfold neural nodes in both time and wavelength and the use of FWM nonlinearity in physics informed neural networks.

**Kostas Sozos** received his B.S degree in Physics from the University of Patras in 2018 and the M.Sc in Microsystems & Nanodevices from the National Technical University of Athens in 2020. He is currently pursuing his Ph.D at the University of West Attica in the field of Neuromorphic Photonic Computing under the supervision of Prof. Adonis Bogris. His main research interests include photonic neuromorphic computing, pattern recognition and optical communications.

**Charis Mesaritakis** received his BS degree in Informatics, from the department of Informatics & Telecommunications of the National & Kapodistrian University of Athens in 2004. He received the MSc in Microelectronics from the same department, whereas in 2011 he received his Ph.D degree on the field of quantum dot devices and systems for next generation optical networks, in the photonics technology & optical communication laboratory of the same institution. In 2012 he was awarded a European scholarship for post-doctoral studies (Marie Curie FP7-PEOPLE IEF) in the joint research facilities of Alcatel-Thales-Lucent in Paris-France, where he worked on intra-satellite communications. He has actively participated as research engineer/technical supervisor in more than 10 EU-funded research programs (FP6-FP7-H2020) targeting excellence in the field of photonic neuromorphic computing, cyber-physical security and photonic integration. He is currently an Associate Professor at the Department of Information & Communication Systems Engineering at the University of the Aegean, Greece. He is the author and co-author of more than 60 papers in highly cited peer reviewed international journals and conferences, two international book chapters, whereas he serves as a regular reviewer for IEEE.

**Adonis Bogris** was born in Athens. He received the B.S. degree in informatics, the M.Sc. degree in telecommunications, and the Ph.D. degree from the National and Kapodistrian University of Athens, Athens, in 1997, 1999, and 2005, respectively. His doctoral thesis was on all-optical processing by means of fiber-based devices. He is currently a Professor at the Department of Informatics and Computer Engineering at the University of West Attica, Greece. He has authored or co-authored more than 150 articles published in international scientific journals and conference proceedings and he has participated in plethora of EU and national research projects. His current research interests include high-speed all-optical transmission systems and networks, nonlinear effects in optical fibers, all-optical signal processing, neuromorphic photonics, mid-infrared photonics and cryptography at the physical layer. Dr. Bogris serves as a reviewer for the journals of the IEEE.